\documentclass[aps,floats,floatfix,showpacs,preprintnumbers,amssymb,prd,twocolumn,superscriptaddress,nofootinbib,nolonBIbliography,reprint]{revtex4-1}

\usepackage[T1]{fontenc}
\usepackage[utf8]{inputenc}               
\usepackage{hyperref}
\hypersetup{
    colorlinks=true,
    linkcolor=blue,
    citecolor=blue,
    urlcolor=blue
}

\usepackage{amssymb,amsmath,verbatim,mathtools,needspace,enumitem,etoolbox,graphicx,physics,microtype,afterpage,xspace,tabularx,lmodern,multirow,boldline, makecell, booktabs}

\usepackage{appendix}
\usepackage{tensor}
\usepackage{array}
\usepackage{siunitx}
\usepackage[normalem]{ulem}
\usepackage[dvipsnames, usenames]{xcolor}
\usepackage{xr-hyper}
\usepackage[font=small,labelfont=bf]{caption}
\definecolor{linkcolor}{rgb}{0.0,0.3,0.5}
\usepackage[all]{hypcap}
\usepackage[T1]{fontenc}
\usepackage[utf8]{inputenc}
\usepackage[usenames,dvipsnames]{xcolor}
\definecolor{cornellGreen}{HTML}{6EB43F}
\definecolor{cornellRed}{HTML}{B31B1B}
\hypersetup{colorlinks=true,citecolor=cornellRed,linkcolor=cornellRed,urlcolor=cornellRed}
\usepackage{multirow}

\definecolor{romared}{RGB}{142,0,28}

\def\bb#1\ee{\begin{equation}\begin{split}#1\end{split}\end{equation}}

\def\[{\left[}
\def\]{\right]}

\def\be{\begin{equation}}
\def\ee{\end{equation}}
\newcommand{\beq}{\begin{eqnarray}}
\newcommand{\eeq}{\end{eqnarray}}

\usepackage{makecell}
\usepackage{soul}

\newcolumntype{Y}{>{\centering\arraybackslash}X}

\usepackage{graphicx}
\usepackage{dcolumn}
\usepackage{bm}
\usepackage{hyperref}
\usepackage{color}

\usepackage{times} 

\usepackage{graphicx} 
\usepackage{booktabs} 
\usepackage[font=small,labelfont=bf]{caption} 
\usepackage{amsfonts, amsmath, amsthm, amssymb} 

\usepackage{wrapfig} 
 \definecolor{mypurple}{RGB}{130, 0, 130} 

\begin{document}
\affiliation{Institute of Fundamental Physics and Quantum Technology, \& School of Physical Science and Technology, Ningbo University, Ningbo, Zhejiang 315211, China}

\title{Exact $T\bar{T}$-Like Flows in Nonlinear Electrodynamics}
\author{Zahra Pezhman}
\email{zpezhmanh@phd.guilan.ac.ir}
\affiliation{Department of Physics, University of Guilan, P.O. Box 41335-1914, Rasht, Iran}

\author{H. Babaei-Aghbolagh}
\email{hosseinbabaei@nbu.edu.cn}
\affiliation{Institute of Fundamental Physics and Quantum Technology, \& School of Physical Science and Technology, Ningbo University, Ningbo, Zhejiang 315211, China}
\author{Komeil Babaei Velni}
\email{babaeivelni@guilan.ac.ir}
\affiliation{Department of Physics, University of Guilan, P.O. Box 41335-1914, Rasht, Iran}

\date{\today}
\begin{abstract}
We derive an exact nonperturbative $T\bar T$-like flow for the class of one-scale, self-dual nonlinear electrodynamics theories admitting a Courant--Hilbert representation. The flow has a universal parametric form in terms of the defining Courant--Hilbert function and, on any physical branch where the relevant invariant map is invertible, becomes an autonomous function of the two quadratic stress-tensor invariants, $\partial_\lambda\mathcal L=X\,\mathcal F(Y/X)$. The function $\mathcal F$ is model dependent, and global autonomy is  assumed when the invariant map is one-to-one. Born--Infeld theory yields a quadratic stress-tensor flow, whereas the causal Russo--Townsend logarithmic model gives an exact nonlinear expression involving the Lambert $W$ function whose weak-field expansion reproduces the known perturbative result. We also establish the equivalence with the Russo--Townsend auxiliary-field formulation. These results provide an exact construction of autonomous irrelevant flows for a broad, explicitly specified class of self-dual nonlinear electrodynamics theories.
\end{abstract}

\maketitle

\section{Introduction}
Fundamental equations can organize broad families of physical theories. The Courant--Hilbert (CH) equation~\cite{10.1115/1.3630089} is one such structure: its solutions describe a large class of self-dual nonlinear electrodynamics (NED) theories and have recently provided a common language for deformation flows in integrable field theories and NED~\cite{Ferko:2023wyi,Ferko:2024zth,Babaei-Aghbolagh:2025uoz,Babaei-Aghbolagh:2025hlm,Fukushima:2025tlj,Fukushima:2026gan}. Related developments include a two-dimensional massive-gravity construction whose on-shell action reproduces the CH deformation~\cite{Fukushima:2026znr,Baglioni:2026mii}.

In stress-tensor flow approaches, a coupling derivative of the Lagrangian is expressed in terms of a function of the duality-invariant energy-momentum tensor of the theory.  Since, a general flow equation takes the form
\begin{equation}\label{fgd}
\frac{\partial \mathcal{L}}{\partial \lambda} = \mathcal{O}_\lambda(T_{\mu\nu}),
\end{equation}
where $\mathcal{O}_\lambda(T_{\mu\nu})$ is an operator constructed from the energy-momentum tensor by two structures $X=   T_{\mu\nu}T^{\mu\nu}$ and $Y=  T_{\mu}{}^{\mu}T_{\nu}{}^{\nu}$.

Born--Infeld (BI) theory is a distinguished example: its deformation from Maxwell theory is governed by a quadratic $T\bar T$-like operator~\cite{Born:1934gh,Babaei-Aghbolagh:2022uij,Babaei-Aghbolagh:2022leo,Ferko:2022cix,Conti:2022egv}. The complementary root-$T\bar T$ structure appears in ModMax theory~\cite{Bandos:2020jsw}. These examples motivate the question of whether an exact stress-tensor flow can be constructed for a broader class of self-dual NED models.

Here we construct such a flow for the one-scale class of CH solutions, defined by $\ell(\tau;\lambda)=\lambda^{-1}f(\lambda\tau)$. We show that every sufficiently differentiable function $f$ in this class determines an exact parametric flow, and that this flow can be written autonomously in terms of $X$ and $Y$ on any physical branch where the corresponding invariant map is invertible. The resulting function $\mathcal F(Y/X)$ depends on the model; the universality lies in the construction and invariant form, not in a single model-independent flow function. This qualification is important because a non-monotonic invariant map may require a branchwise description.

BI theory provides a closed-form benchmark for which the flow reduces to a quadratic stress-tensor operator. The causal Russo--Townsend logarithmic model~\cite{Russo:2024kto} supplies a genuinely nonlinear example: its invariant map can be inverted using the lower real branch of the Lambert $W$ function, yielding an exact flow whose weak-field expansion agrees with the previously derived perturbative result~\cite{Babaei-Aghbolagh:2025cni}. We further show that the same construction follows from the Russo--Townsend auxiliary-field formulation~\cite{Russo:2025fuc}, after consistently matching its potential to the CH function. The new result is therefore the exact parametric construction and its branchwise autonomous representation, namely that the $T\bar{T}$-like flow equation for a generic NED can be cast into a form with no explicit dependence on the coupling parameter.

The paper is organized as follows. We review the CH representation and derive the general flow, then examine BI and logarithmic examples. We finally formulate the result in auxiliary-field variables and summarize the scope and limitations of the construction.

\section{Courant--Hilbert approach}
We consider local electromagnetic theories whose Lagrangian depends on the field strength $F_{\mu\nu}$ but not on its derivatives. We use the conventions
\begin{equation}
S=-\frac14F_{\mu\nu}F^{\mu\nu},\qquad
P=-\frac14F_{\mu\nu}\widetilde F^{\mu\nu},
\end{equation}
with $\widetilde F^{\mu\nu}=\tfrac12\epsilon^{\mu\nu\rho\sigma}F_{\rho\sigma}$, so that the Maxwell Lagrangian is $\mathcal L=S$. Equivalently, one may use the non-negative combinations
\begin{equation}
U=\frac{\sqrt{S^2+P^2}-S}{2},\qquad
V=\frac{\sqrt{S^2+P^2}+S}{2}.
\label{eq:UV}
\end{equation}
In these conventions, electromagnetic self-duality is encoded by the Courant--Hilbert (CH) equation~\cite{Gaillard:1997rt,Gaillard:1997zr,Gibbons:1995cv,Sorokin:2021tge}
\begin{equation}
\mathcal L_U\mathcal L_V=-1,
\label{CHEQ}
\end{equation}
where subscripts denote partial derivatives. Its general solution can be written implicitly in terms of a single function $\ell(\tau)$:
\begin{equation}
\mathcal L=\ell(\tau)-\frac{2U}{\dot\ell(\tau)},\qquad
\tau=V+\frac{U}{\dot\ell(\tau)^2},
\label{Lg}
\end{equation}
where $\dot\ell=d\ell/d\tau\ne0$. The physical domain has $\tau\geq0$, with $\tau=0$ at the vacuum $U=V=0$.

The energy-momentum tensor of a CH solution decomposes as~\cite{Russo:2024xnh,Russo:2024ptw}
\begin{equation}
T_{\mu\nu}=\frac{\tau\dot\ell}{U+V}T_{\mu\nu}^{\mathrm{Max}}+(\ell-\tau\dot\ell)g_{\mu\nu},
\label{Tmnu}
\end{equation}
where $T_{\mu\nu}^{\mathrm{Max}}$ is traceless. Thus
\begin{equation}
T^\mu{}_{\mu}=4(\ell-\tau\dot\ell),
\end{equation}
and the quadratic invariants are
\begin{equation}
X=4\left[(\tau\dot\ell)^2+(\ell-\tau\dot\ell)^2\right],\qquad
Y=16(\ell-\tau\dot\ell)^2.
\label{TT}
\end{equation}
These expressions will allow us to construct the flow directly from the CH data.

\section{Exact $T\bar T$-like flow}
We focus on the one-scale class
\begin{equation}
z=\lambda\tau,\qquad
\ell(\tau;\lambda)=\frac{1}{\lambda}f(z),
\label{one_scale}
\end{equation}
where $[\lambda]=-4$ and $[\ell]=[\tau]=4$, so that $z$ is dimensionless. We do not assume that $f$ is analytic; differentiability sufficient to define the displayed expressions is understood. The Maxwell weak-field normalization is $f(0)=0$ and $f'(0)=1$, so $\ell=\tau+o(\tau)$ as $\lambda\tau\to0$. When a Taylor expansion exists, this can be sharpened to the usual power-series expansion in $\lambda$.

Define the dimensionless functions
\begin{equation}
a(z)=zf'(z),\qquad b(z)=f(z)-zf'(z).
\label{ab2}
\end{equation}
Then
\begin{equation}
\tau\dot\ell=\frac{a(z)}{\lambda},\qquad
\ell-\tau\dot\ell=\frac{b(z)}{\lambda},
\label{ab}
\end{equation}
and the stress-tensor invariants become
\begin{equation}
X=\frac{4[a(z)^2+b(z)^2]}{\lambda^2},\qquad
Y=\frac{16b(z)^2}{\lambda^2}.
\label{TTab}
\end{equation}
Their dimensionless ratio is therefore
\begin{equation}
\mathcal R(z)\equiv\frac{Y}{X}=\frac{4b(z)^2}{a(z)^2+b(z)^2}.
\label{YoverX}
\end{equation}

At fixed electromagnetic invariants, the CH parametrization implies
\begin{equation}
\left.\frac{\partial\mathcal L}{\partial\lambda}\right|_{U,V}
=\left.\frac{\partial\ell}{\partial\lambda}\right|_{\tau}
=-\frac{b(z)}{\lambda^2}.
\label{flow_derivative}
\end{equation}
Combining this identity with \eqref{TTab} gives the exact parametric flow
\begin{equation}
\boxed{\frac{\partial\mathcal L}{\partial\lambda}=X F(z),\qquad
F(z)=-\frac{b(z)}{4[a(z)^2+b(z)^2]}.}
\label{f2}
\end{equation}
This expression is valid wherever the one-scale representation is regular and $a^2+b^2\ne0$. At the vacuum, where $X=Y=0$, the flow is understood by its limit from nonzero fields. If $f$ is twice differentiable at $z=0$, then $F(0)=f''(0)/8$.

The parametric flow becomes an autonomous function of the stress-tensor invariants if $\mathcal R(z)$ is one-to-one on the physical branch being considered. Writing its inverse as $z=Z(\mathcal R)$ and defining $\mathcal F(\mathcal R)=F(Z(\mathcal R))$, we obtain
\begin{equation}
\boxed{\frac{\partial\mathcal L}{\partial\lambda}=X\,\mathcal F\!\left(\frac{Y}{X}\right).}
\label{gold}
\end{equation}
The function $\mathcal F$ is generally model dependent. The universal statement is that the one-scale CH construction yields this invariant form whenever the inverse exists; if $\mathcal R(z)$ is not globally one-to-one, the flow is defined branchwise. The right-hand side then has no explicit dependence on $\lambda$, $\tau$, $U$, or $V$.

For some models it is useful to retain the sign information lost in $Y$. On a physical branch with $a(z)/\lambda=\tau\dot\ell>0$, define
\begin{equation}
r(z)\equiv-\frac{b(z)}{a(z)}=-\frac{T^\mu{}_{\mu}}{\sqrt{4X-Y}}.
\label{rdef}
\end{equation}
It follows that
\begin{equation}
\frac{Y}{X}=\frac{4r^2}{1+r^2},\qquad
r=-\operatorname{sgn}(T^\mu{}_{\mu})\frac{\rho}{\sqrt{4-\rho^2}},
\qquad \rho\equiv\sqrt{\frac{Y}{X}}.
\label{r}
\end{equation}
Here the invariant ratios are defined away from the vacuum and extended to it only by limits. If $r(z)$ is invertible on the selected physical branch, the auxiliary parameter can equivalently be eliminated using $z=z(r)$; otherwise the result remains parametric or branchwise.

\section{Born--Infeld theory}
Born--Infeld (BI) theory is a standard example of a causal, duality-invariant nonlinear electrodynamics and regularizes the electrostatic self-energy of a point charge~\cite{Born:1934gh}. Its CH function is~\cite{Russo:2024llm}
\begin{equation}
\ell_{\mathrm{BI}}(\tau;\lambda)=\frac{1}{\lambda}\left(1-\sqrt{1-2\lambda\tau}\right).
\label{CHBI}
\end{equation}
Substitution into \eqref{Lg} gives
\begin{equation}
\mathcal L_{\mathrm{BI}}=\frac{1}{\lambda}\left[1-\sqrt{(1+2\lambda U)(1-2\lambda V)}\right].
\label{MMB}
\end{equation}
In the one-scale notation, $f(z)=1-\sqrt{1-2z}$. Setting $s=\sqrt{1-2z}$, the functions in \eqref{ab2} are
\begin{equation}
a=\frac{1-s^2}{2s},\qquad b=-\frac{(1-s)^2}{2s},
\end{equation}
which give
\begin{equation}
F_{\mathrm{BI}}(s)=\frac{s}{4(1+s^2)}.
\end{equation}
The signed ratio $r=-b/a$ is
\begin{equation}
r=\frac{1-s}{1+s},\qquad s=\frac{1-r}{1+r},
\end{equation}
so that
\begin{equation}
\mathcal F_{\mathrm{BI}}(r)=\frac{1-r^2}{8(1+r^2)}.
\end{equation}
Using $r^2/(1+r^2)=Y/(4X)$, the flow is exactly
\begin{equation}
\boxed{\frac{\partial\mathcal L_{\mathrm{BI}}}{\partial\lambda}
=\frac18X-\frac1{16}Y
=\frac18\left[T_{\mu\nu}T^{\mu\nu}-\frac12(T^\mu{}_{\mu})^2\right].}
\label{ttbi}
\end{equation}
Thus BI theory provides a closed-form example in which the autonomous flow is quadratic in the stress tensor. The physical CH domain for $\lambda>0$ is $0\le z\le1/2$, or equivalently $0\le s\le1$ and $0\le r\le1$. The endpoints are $z=0,s=1,r=0$ at the vacuum and $z=1/2,s=0,r=1$ at the maximal-$\tau$ endpoint. The map is monotonic on this interval, so the autonomous representation is single valued on the full physical branch.

We now compare this polynomial flow with a causal model whose exact flow is genuinely nonlinear.

\section{Exact flow of the logarithmic theory}
\label{sec:log}
The causal logarithmic model of Russo and Townsend~\cite{Russo:2024kto} is defined by the CH function
\begin{equation}
\ell_{\log}(\tau;\lambda)=-\frac{1}{\lambda}\log(1-\lambda\tau),
\quad f(z)=-\log(1-z).
\label{chlog}
\end{equation}
Substituting this function into \eqref{Lg} gives the corresponding self-dual logarithmic Lagrangian,
\begin{eqnarray}\label{Log}
	\mathcal{L}_{Log}& =&\frac{1}{\lambda} \biggl[1 -  \sqrt{1 + 4  \lambda U  - 4  \lambda^2 U  \, V}\\
    &&-  \log\biggl(\frac{\sqrt{1 + 4  \lambda U  - 4  \lambda^2 U  \, V}-1}{2  \lambda U}\biggr) \biggr]\,,\nonumber
\end{eqnarray}
where the expression is understood by continuity when $U\to0$. Equivalently, the implicit CH representation \eqref{Lg} may be used without introducing this explicit form.

The functions controlling the flow are
\begin{equation}
a(z)=\frac{z}{1-z},\qquad
b(z)=-\frac{z}{1-z}-\log(1-z),
\label{logabb}
\end{equation}
and hence
\begin{equation}
r(z)=-\frac{b(z)}{a(z)}=1+\frac{1-z}{z}\log(1-z).
\label{rlog}
\end{equation}
Solving for $z$ gives
\begin{equation}
z(r)=1-\frac{r-1}{W_{-1}\!\left((r-1)e^{r-1}\right)},
\label{zlog}
\end{equation}
where $W_{-1}$ is the lower real branch of the Lambert $W$ function. On the nontrivial physical interval $0<z<1$, one has $0<r<1$; the $W_{-1}$ branch selects the nontrivial inverse. The principal branch $W_0$ gives the degenerate solution $z=0$ and is relevant only to the vacuum limit. The endpoint limits should therefore be taken continuously rather than by substituting directly into the inverse formula.

Using the general expression $F=r/[4a(1+r^2)]$, the exact flow becomes
\begin{equation}
\boxed{\frac{\partial\mathcal L_{\log}}{\partial\lambda}
=X\,\mathcal F_{\log}(\rho),\qquad
\mathcal F_{\log}(\rho)=\frac{r[1-z(r)]}{4z(r)[1+r^2]},}
\label{nonLog}
\end{equation}
where $r=\rho/\sqrt{4-\rho^2}$ on this branch and $\rho=\sqrt{Y/X}$. Expanding about the weak-field point yields
\begin{equation}
\begin{aligned}
\mathcal F_{\log}(\rho)={}&\frac18-\frac{\rho}{12}-\frac{7\rho^2}{288}
+\frac{49\rho^3}{4320}+\frac{\rho^4}{6480}
+\mathcal O(\rho^5).
\end{aligned}
\label{Flogexp}
\end{equation}
This agrees with the perturbative logarithmic flow found in~\cite{Babaei-Aghbolagh:2025cni}. In contrast to BI theory, the odd powers of $\rho$ are nonzero, producing half-integer powers of the quadratic stress-tensor invariants in the operator expansion. The physical branch has $\lambda>0$, $0<z<1$, $0<r<1$, and $0<\rho<\sqrt2$; the vacuum is approached as a limiting endpoint.

\section{Auxiliary-field formulation}
The Russo--Townsend formulation provides an alternative description of self-dual NED in terms of an auxiliary field $\varphi$ and a potential $W(\varphi)$~\cite{Russo:2024ptw,Russo:2025fuc}:
\begin{equation}
\mathcal L(U,V;\varphi)=-e^{-\varphi}U+e^{\varphi}V-W(\varphi).
\label{lsp}
\end{equation}
The auxiliary equation of motion, $\partial\mathcal L/\partial\varphi=0$, enforces the self-duality relation. Matching this formulation to the CH solution gives the Legendre-transform dictionary
\begin{equation}
\boxed{W(\varphi)=\tau\dot\ell-\ell,\qquad
W'(\varphi)=\tau\dot\ell,\qquad e^{\varphi}=\dot\ell(\tau).}
\label{dict}
\end{equation}
Here $W'=dW/d\varphi$. In particular, the sign in $W=\tau\dot\ell-\ell$ is fixed by the minus sign in the auxiliary Lagrangian \eqref{lsp}; it ensures that the metric term in the stress tensor agrees with \eqref{Tmnu}.

The auxiliary-field stress tensor is
\begin{equation}
T_{\mu\nu}=\frac{W'(\varphi)}{U+V}T_{\mu\nu}^{\mathrm{Max}}-W(\varphi)g_{\mu\nu},
\label{Tmnuphi}
\end{equation}
so that
\begin{equation}
X=4\left[(W')^2+W^2\right],\qquad Y=16W^2.
\label{XYW}
\end{equation}
These expressions agree with the CH invariants because $W=-b/\lambda$ and $W'=a/\lambda$. Define $w(\varphi)=\lambda W(\varphi)$, for $\lambda>0$. 
It seems that the free-theory limit, $\lambda \to 0$, is not clearly defined; equivalently, $W(\varphi)$ is not defined for Maxwell's theory. An alternative auxiliary field formulation was given in~\cite{Kuzenko:2025gvn,Kuzenko:2026hbw}, in which ModMax can be obtained but Maxwell cannot. However, we take the seed theory (Maxwell's theory) as the basis of the well-defined main theory, independently of the $T\bar{T}$ flow equation already given in~\eqref{fgd}.
The exact flow is then
\begin{equation}
\frac{\partial\mathcal L}{\partial\lambda}=\frac{w}{\lambda^2}
=X \mathfrak{F}(\varphi),\qquad
\mathfrak{F}(\varphi)=\frac{w(\varphi)}{4\left[(\partial_\varphi w)^2+w(\varphi)^2\right]}.
\label{ffr}
\end{equation}
The notation $\partial_\varphi w$ denotes the derivative with respect to the auxiliary field. On a branch where $w$ is locally invertible in the signed ratio $r=w/(\partial_\varphi w)$, one can reconstruct the auxiliary field from the flow function. In particular,
\begin{equation}
w(\varphi(r))=\frac{r^2}{4(1+r^2)\mathfrak{F}(r)},
\end{equation}
and therefore
\begin{equation}
\varphi(r)=2\arctan r-\int_0^r t\,\frac{\mathfrak{F}'(t)}{\mathfrak{F}(t)}\,dt,
\label{ffdw}
\end{equation}
where $\mathfrak{F}'(t)=d\mathfrak{F}/dt$. This representation reproduces the CH flow when the same physical branch and sign conventions are used. It also makes clear that the equivalence is branchwise whenever the relevant inverse map is not globally one-to-one.

\section{Conclusion and outlook}

We have derived an exact nonperturbative $T\bar T$-like flow for the one-scale class of self-dual nonlinear electrodynamics theories admitting the CH representation $\ell(\tau;\lambda)=\lambda^{-1}f(\lambda\tau)$. The flow always has a parametric form in terms of the CH data and becomes an autonomous function of the quadratic stress-tensor invariants when the associated invariant map is invertible on the chosen physical branch. The function specifying the autonomous flow is model dependent; the universal result is the construction and its invariant structure. Born--Infeld theory yields a quadratic stress-tensor operator, whereas the causal logarithmic model gives an exact Lambert-$W$ representation with a nontrivial weak-field expansion. The corrected Legendre dictionary also establishes agreement with the Russo--Townsend auxiliary-field formulation.

Several questions remain. It would be useful to determine which other CH models admit globally single-valued autonomous flows, whether the CH function can be reconstructed from $\mathcal F(Y/X)$, and how causality and $\varphi$-parity constrain the resulting flow functions. Extensions to multiple couplings and to other dimensions are further directions, but require separate analysis and are not assumed in the present derivation.

\begin{acknowledgments}
We are grateful to Christian Ferko, Song He, Stefano Negro, Jorge G. Russo, Dmitri Sorokin, and Roberto Tateo for their interest in this work and the fruitful discussions that followed. We would like to express our special thanks to Dmitri Sorokin for his valuable insights, generous feedback, and encouraging discussions throughout this project. The work of H.B.-A. was conducted as part of the PostDoc Program on {\it Exploring TT-bar Deformations: Quantum Field Theory and Applications,} sponsored by Ningbo University.
\end{acknowledgments}

\bibliography{Ref}

\end{document}